\documentclass[aps,prl,twocolumn,showpacs,byrevtex]{revtex4}

\usepackage{times,mathptm}% for making `pdf' file
\usepackage[dvips]{graphicx,color}% for inclusion of graphics file(s)
\usepackage{titlesec}
\newcommand{\bqa}{\begin{eqnarray*}}
\newcommand{\eqa}{\end{eqnarray*}}

\begin{document}

\title{Origin of High-Temperature Superconductivity in Compressed LaH$_{10}$}
\author{Liangliang Liu$^{1,2}$, Chongze Wang$^1$, Seho Yi$^1$, Kun Woo Kim$^3$, Jaeyong Kim$^1$, and Jun-Hyung Cho$^{1*}$}
\affiliation{
$^1$Department of Physics, Research Institute for Natural Science, and HYU-HPSTAR-CIS High Pressure Research Center, Hanyang
University, 222 Wangsimni-ro, Seongdong-Ku, Seoul 04763, Korea \\
$^2$ Key Laboratory for Special Functional Materials of Ministry of Education, Henan University, Kaifeng 475004, People's Republic of China  \\
$^3$ Center for Theoretical Physics of Complex Systems, Institute for Basic Science, Daejeon 34051, Republic of Korea}

%\email{chojh@hanyang.ac.kr}
\date{\today}

\begin{abstract}
Room-temperature superconductivity has been one of the most challenging subjects in modern physics. Recent experiments reported that lanthanum hydride LaH$_{10{\pm}x}$ ($x$$<$1) raises a superconducting transition temperature $T_{\rm c}$ up to ${\sim}$260 (or 250) K at high pressures around 190 (170) GPa. Here, based on first-principles calculations, we reveal that compressed LaH$_{10}$ has the topological Dirac-nodal-line (DNL) states that create a van Hove singularity (vHs) near the Fermi energy. Contrasting with the previously proposed bonding nature based on a charge transfer picture from cationic La to anionic H, we identify a peculiar characteristic of electrical charges with anionic La and both cationic and anionic H species, caused by a strong hybridization of the La $f$ and H $s$ orbitals. We find that a large number of electronic states at the vHs are strongly coupled to the H-derived high-frequency phonon modes that are induced via the unusual, intricate bonding network of LaH$_{10}$, thereby yielding a high $T_{\rm c}$. Our findings elucidate the microscopic origin of the observed high-$T_{\rm c}$ BCS-type superconductivity in LaH$_{10}$, which can be generic to another recently observed high-$T_{\rm c}$ hydride H$_3$S.
\end{abstract}
%\pacs{75.10.Lp, 75.30.Hx, 75.30.Et}

%\begin{document}
\maketitle

%\begin{multicols}{2}
%\vspace{0.4cm}
%\section{INTRODUCTION}
%\vspace{0.4cm}

Ever since the first discovery of superconductivity (SC) in 1911~\cite{historyCon}, scientists have searched for materials that can conduct electricity without resistance below a superconducting transition temperature $T_{\rm c}$. So far, tremendous efforts have been devoted to exploring high-$T_{\rm c}$ SC in a variety of materials such as cuprates~\cite{YBaCuO}, pnictides~\cite{Fesuper1}, and hydrogen-rich compounds (called hydrides)~\cite{Hydride1,Hydride4,Hydride5,Hydride7}. Among them, SC in hydrides has been successfully explained by the Bardeen-Cooper-Schrieffer (BCS) theory~\cite{BCS}, where SC is driven by a condensate of electron pairs, so-called Cooper pairs, due to electron-phonon interaction. The pioneering idea that hydrogen can be a good candidate for high-$T_{\rm c}$ SC was proposed by Ashcroft~\cite{Ashc}: i.e., metallic hydrogen with light atomic mass is expected to have high vibrational frequencies, thereby providing a high $T_{\rm c}$ due to a strong electron-phonon coupling (EPC). However, the metallization of hydrogen is very difficult to achieve experimentally, because it requires too high pressures over ${\sim}$400 GPa ~\cite{MetalicH1,MetalicH2,MetalicH3}. Instead, high-$T_{\rm c}$ SC in hydrides can be realized at relatively lower pressures that are currently accessible using static compression techniques~\cite{Hydride8}. Motivated by theoretical predictions of high-$T_{\rm c}$ SC in many hydrides~\cite{Hydride5,Hydride9,Hydride11}, experiments were performed to confirm that compressed sulfur hydride H$_3$S exhibits a $T_{\rm c}$ of 203 K at pressures around 150 GPa~\cite{ExpH3S}. Recently, X-ray diffraction and optical studies demonstrated that lanthanum (La) hydrides can be synthesized in an $fcc$ lattice at ${\sim}$170 GPa upon heating to ${\sim}$1000 K~\cite{LaH}, consistent with the earlier predicted cubic metallic phase of LaH$_{10}$ having cages of thirty-two H atoms surrounding an La atom [see Fig. 1(a)]~\cite{Hydride5,Hydride7}. Subsequently, two experimental groups nearly simultaneously reported that such an La hydride LaH$_{10{\pm}x}$ ($x$ $<$ 1)~\cite{Note1} exhibits a high $T_{\rm c}$ ${\approx}$ 260 K at pressures of ${\sim}$190 GPa~\cite{LaH2} or 250 K at ${\sim}$170 GPa ~\cite{LaH3,LaH4}.

To explore the origin of the high-temperature BCS-type SC of compressed LaH$_{10}$, we here, based on first-principles calculations, not only investigate the salient characteristics of the electronic, bonding, and phononic properties, but also identify which electronic states are coupled with H-derived high-frequency phonon modes. We reveal that LaH$_{10}$ has the topological Dirac-nodal-line (DNL) states including seven one-dimensional (1D) dispersive nodal lines near the Fermi energy $E_{\rm F}$. Remarkably, four DNLs are split into hole-like and electron-like bands at the high-symmetry point $L$ on the Brillouin zone boundary, forming a van Hove singularity (vHs) close to $E_{\rm F}$. We find that the electronic states near $E_{\rm F}$ exhibit a strong hybridization of the La 4$f$ and H $s$ orbitals, leading to a peculiar electrical charge characteristic of anionic La and both cationic and anionic H atoms. This unusual, intricate bonding network of LaH$_{10}$ induces the high-frequency vibrations of H atoms, which are, in turn, strongly coupled to large numbers of electronic states at the vHs. The resulting strong EPC significantly enhances $T_{\rm c}$, as measured by recent experiments~\cite{LaH2,LaH3,LaH4}. Therefore, our findings shed light on the microscopic origin of the high-$T_{\rm c}$ BCS-type SC observed in compressed LaH$_{10}$.

\begin{figure}[htb]
\centering{ \includegraphics[width=8.5cm]{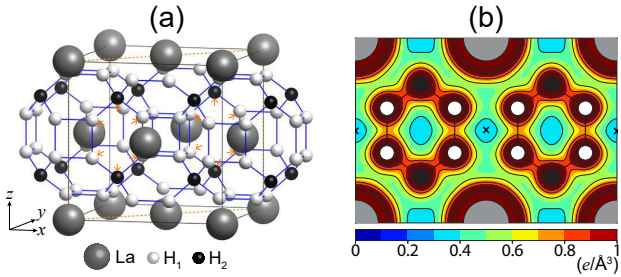} }
\caption{(Color online) (a) Optimized structure of the $fcc$ LaH$_{10}$ crystal. The dashed line represents the (1$\bar{1}$0) face. The H atoms located on the (1$\bar{1}$0) face are indicated by the arrows. (b) Total charge density ${\rho}_{\rm tot}$ of LaH$_{10}$, plotted on the (1$\bar{1}$0) face. In (b), the contour spacing is 0.2 $e$/{\AA}$^3$. The void space surrounded by eight neighboring H$_1$ atoms is marked ${\times}$ in (b). }
\end{figure}

We have performed the density-functional theory (DFT) calculations for the electronic, bonding, and phononic properties of compressed LaH$_{10}$ at 300 GPa. We used the Vienna {\it ab initio} simulation package with the projector-augmented wave method~\cite{vasp1,vasp2,paw}. For the exchange-correlation energy, we employed the generalized-gradient approximation functional of Perdew-Burke-Ernzerhof (PBE)~\cite{pbe}. A plane-wave basis was taken with a kinetic energy cutoff of 500 eV. The ${\bf k}$-space integration was done with 16${\times}$16${\times}$16 $k$ points (in the Brillouin zone) for the structure optimization and 70${\times}$70${\times}$70 $k$ points for the density of states (DOS) calculation. All atoms were allowed to relax along the calculated forces until all the residual force components were less than 0.005 eV/{\AA}. For the computations of the phonons and EPC constants, we used the QUANTUM ESPRESSO package~\cite{QE} to employ the 6${\times}$6${\times}$6 $q$ points and 24${\times}$24${\times}$24 $k$ points, respectively.

We begin by optimizing the structure of compressed LaH$_{10}$ at 300 GPa using the DFT calculation. Figure 1(a) shows the optimized $fcc$ LaH$_{10}$ structure, which has the lattice parameters $a$ = $b$ = $c$ = 4.748 {\AA}. We find that the H$_{1}-$H$_1$ bond length $d_1$ is 1.145 {\AA}, slightly longer than the H$_{1}-$H$_2$ bond length $d_2$ = 1.064 {\AA}. These values are in good agreement with those ($d_1$ = 1.152 {\AA} and $d_2$ = 1.071 {\AA}) of a previous DFT calculation~\cite{Hydride5} performed at 300 GPa. The calculated total charge density ${\rho}_{\rm tot}$ of LaH$_{10}$ is displayed in Fig. 1(b). It is seen that H atoms in the H$_{32}$ cage are bonded to each other with weakly covalent bonds. Here, each H-H bond has a saddle point of charge density at its midpoint, similar to the C-C covalent bond in diamond~\cite{diamond}. The charge densities at the midpoints of the H$_{1}-$H$_1$ and H$_{1}-$H$_2$ bonds are 0.74 and 0.91 e/{\AA}$^3$, respectively. The relatively weaker H$_{1}-$H$_1$ covalent bond is well represented by its larger value of $d_1$ than d$_2$. On the other hand, the electrical charges of La and H atoms are well separated, indicating an ionic character. We note that for other La hydrides such as LaH$_{2+x}$ with fluorite-type structures~\cite{LnH2}, a charge transfer occurs from La to H atoms, thereby resulting in cationic La and anionic H. But surprisingly, compressed LaH$_{10}$ is found to exhibit a drastically different feature of electrical charges, i.e., anionic La and both anionic H$_1$ and cationic H$_2$, as discussed below. This unusual, intricate bonding network of LaH$_{10}$ contrasts with the previously proposed bonding nature based on a charge transfer picture from cationic La to anionic H~\cite{Hydride7}.

\begin{figure}[ht]
\includegraphics[width=8.5cm]{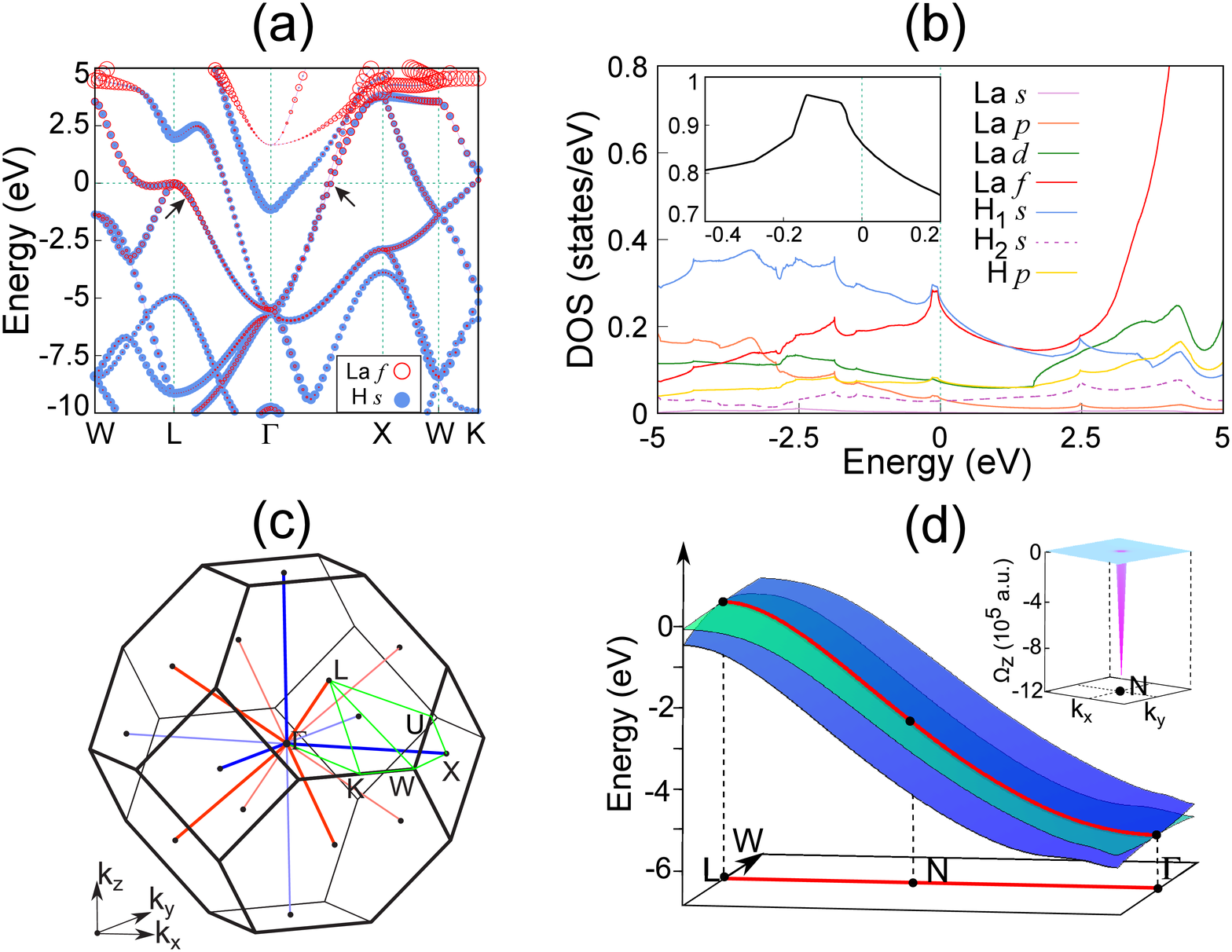}
\caption{(Color online) (a) Calculated band structure and (b) partial DOS of LaH$_{10}$. In (a), the bands projected onto the H $s$ and La $f$ orbitals are displayed with circles whose radii are proportional to the weights of the corresponding orbitals. The arrows in (a) indicate the DNLs along the $\overline{{\Gamma}L}$ and $\overline{{\Gamma}X}$ lines. The energy zero represents $E_{\rm F}$. A close-up of the total DOS around the vHs is given in the inset of (b). In (c), all the DNLs along the symmetry-equivalent $\overline{{\Gamma}L}$ and $\overline{{\Gamma}X}$ lines are drawn in the first Brillouin zone. The 1D dispersion of the DNL bands along the $\overline{{\Gamma}L}$ line is displayed in (d), together with the distribution of a Berry curvature component ${\Omega}_{z}$ around a band touching point $N$ in the DNL.}
\end{figure}

Figures 2(a) and 2(b) show the calculated band structure and partial DOS (PDOS) of LaH$_{10}$, respectively. The band projections onto the H $s$ and La $f$ orbitals are also displayed in Fig. 2(a). Compared to other orbitals, these two orbitals are more dominant components of the electronic states located near $E_{\rm F}$ (see Fig. S1 of the Supplemental Material~\cite{SM}). In Fig. 2(a), the band dispersion around the $L$ point (just below $E_{\rm F}$) shows the presence of hole-like and electron-like bands along the $\overline{LW}$ symmetry line, giving rise to an equal vHs around the four symmetry-equivalent $L$ points [see Fig. 2(c)] in the first Brillouin zone. Using the local principal-axis coordinates, we obtain the effective masses $m_1$, $m_2$, and $m_3$ of the hole-like band as $-$0.16$m_e$, $-$1.05$m_e$, and $-$6.95$m_e$ (thermal mass $m_{\rm th}$ ${\equiv}$ $|$$m_1$$m_2$$m_3$$|$$^{1/3}$ = 1.06$m_e$), while we obtain those of electron-like band as $-$0.09$m_e$, 0.79$m_e$, and 0.20$m_e$ ($m_{\rm th}$ = 0.24$m_e$). Note that the larger the thermal mass, the more flatness the hole- or electron-like band. The inset of Fig. 2(b) displays a close-up of the total DOS around the vHs, which represents the presence of two van Hove singularities separated by ${\Delta}E_{\rm vHs}$ = ${\sim}$90 meV. Recently, Quan and Pickett~\cite{thermal} pointed out the importance of such a double-shaped vHs in increasing $T_{\rm c}$ of compressed H$_3$S, where the values of $m_{\rm th}$ are 0.42$m_e$ and 0.31$m_e$ at the two vHs points (with ${\Delta}E_{\rm vHs}$ = ${\sim}$300 meV), respectively. We note that formation of the vHs is associated with symmetry-guaranteed DNLs (see Fig. S2 of the Supplemental Material~\cite{SM}), as discussed below. Thus, the two recently observed high-$T_{\rm c}$ hydrides, H$_3$S~\cite{ExpH3S} and LaH$_{10}$~\cite{LaH2,LaH3}, have similar electronic features such as DNLs and double-shaped vHs, which can be rather generic to high-$T_{\rm c}$ superconductivity. Further, we notice that high-symmetry metallic hydrides having such vHs with large $m_{\rm th}$ can increase their electronic density of states in proportion to the number of symmetry-equivalent vHs points, which results in enhancing $T_{\rm c}$. Interestingly, the PDOS projected onto the H$_1$ $s$ and La $f$ orbitals exhibits sharp peaks close to $E_{\rm F}$ [see Fig. 2(b)], indicating a strong hybridization of the two orbitals. Compared to the H$_1$ $s$ orbital, the PDOS projected onto the H$_2$ $s$ orbital is much suppressed around $E_{\rm F}$. Such different aspects of the H$_1$ and H$_2$ $s$ orbitals reflect their opposite charge characters, i.e., anionic H$_1$ and cationic H$_2$, which will be manifested below by charge-density analysis.

It is remarkable that the hole-like and electron-like bands are degenerate at the $L$ point to form a DNL along the $\overline{{\Gamma}L}$ line, showing the 1D nodal line with a large band width ${\sim}$6 eV [see Fig. 2(a)]. Here, we emphasize that the DNL bands crossing the $L$ point create double-shaped vHs close to $E_{\rm F}$. Figure 2(c) displays all the DNLs near $E_{\rm F}$, which are computed by using the WannierTools package~\cite{wanniertool}. Here, the tight-binding Hamiltonian with maximally localized Wannier functions~\cite{wannier90} reproduces well the electronic bands obtained using the DFT calculation (see Fig. S3 of the Supplemental Material~\cite{SM}). Since the crystalline symmetry of LaH$_{10}$ belongs to the space group $Fm$-$\overline{3}m$ (No. 225) with the point group O$_h$, there are four and three DNLs along the symmetry-equivalent $\overline{{\Gamma}L}$ and $\overline{{\Gamma}X}$ lines [see Fig. 2(c)], respectively. It is noted that along the $\overline{{\Gamma}L}$ line, the system has three-fold rotation symmetry ($C_{3}$) and an inversion ($P$). These spatial symmetries and time ($T$) reversal symmetry~\cite{NLsymm,Kawakami} allow two-dimensional irreducible symmetry representations to protect the DNL along the $\overline{{\Gamma}L}$ line. Similarly, the DNL along the $\overline{{\Gamma}X}$ line is protected by $C_{4}$ and $PT$. As shown in Fig. 2(d), we find that the band touching points along the DNLs exhibit large Berry curvature distributions. To find the topological characterization of the DNLs, we calculate the topological $Z_2$ index~\cite{Z2index}, defined as ${\zeta}_1$ = ${\frac{1}{\pi}}$ ${\oint}$$_c$ $dk$${\cdot}$A($k$), along a closed loop encircling any of the DNLs. Here, A(k) = -$i$$<$$u_k$$\mid$$\partial$$_k$$\mid$$u_k$$>$ is the Berry connection of the related Bloch bands. We obtain ${\zeta}_1$ = 1 for the DNLs along the $\overline{{\Gamma}L}$ and $\overline{{\Gamma}X}$ lines, indicating that they are stable against perturbations without breaking rotational and $PT$ symmetries. Nevertheless, we find that, when the spin-orbit coupling (SOC) is considered, the DNLs are fully gapped with small gaps~\cite{Shao} (see Fig. S4 of the Supplemental Material~\cite{SM}). Although the largest gap (${\sim}$125 meV) is located at the $L$ point, the total DOS around the vHs is similar to that obtained from the DFT calculation without including SOC (see Fig. S5 of the Supplemental Material~\cite{SM}), which may therefore hardly affect $T_{\rm c}$ of LaH$_{10}$.

In order to examine the ionic character between the La atoms and the H$_{32}$ cages, we calculate the charge density difference, defined as ${\Delta}{\rho}$ = ${\rho}_{\rm tot}$ $-$ ${\rho}_{\rm La}$ $-$ ${\rho}_{\rm H}$, where the second and third terms represent the charge densities of the LaH$_0$ lattice (i.e., LaH$_{10}$ structure without H atoms) and the La$_0$H$_{10}$ lattice (LaH$_{10}$ structure without La atoms), respectively. Figures 3(a), 3(b), and 3(c) show ${\rho}_{\rm La}$, ${\rho}_{\rm H}$, and ${\Delta}{\rho}$, respectively. Obviously, ${\Delta}{\rho}$ illustrates that electron charge is transferred from the H$_2$ atoms as well as the void space (marked '${\times}$' region in Fig. 1(b), surrounded by eight neighboring H$_1$ atoms) to the La and H$_1$ atoms. It is thus likely that the La and H$_1$ atoms are characterized as being anionic, while the H$_2$ atoms as being cationic. Such charge characters of LaH$_{10}$ contrast with other La hydrides~\cite{LnH2} that comprise cationic La and anionic H atoms. This drastic difference between LaH$_{10}$ and other La hydrides can be ascribed to their different locations of H $s$ orbitals. For LaH$_{2+x}$ with fluorite-type structures~\cite{LnH2}, the H $s$ orbital is located deeper than the La valence orbitals, therefore inducing a charge transfer from La to H atoms. Meanwhile, as shown in Figs. 2(b) and 3(e), the PDOS of LaH$_{10}$ and the La$_0$H$_{10}$ lattice indicates that near $E_{\rm F}$, the $s$ orbital of anionic H$_1$ is more dominant than that of cationic H$_2$. Consequently, in LaH$_{10}$, the former $s$ orbital hybridizes quite strongly with the La $f$ orbitals to produce their PDOS peaks [see Fig. 2(b)]. Because of the unusual, intricate bonding nature of anionic La/H$_1$ and cationic H$_2$ in LaH$_{10}$, it is expected that a long-ranged Coulomb interaction between constituent atoms induces high-frequency phonon modes, as demonstrated below. Indeed, we find that the frequencies of the ${\Gamma}$-phonon modes in the La$_0$H$_{10}$ lattice are much reduced, compared to those in LaH$_{10}$ (see Fig. S6 of the Supplemental Material~\cite{SM}).

\begin{figure}[htb]
\includegraphics[width=8.5cm]{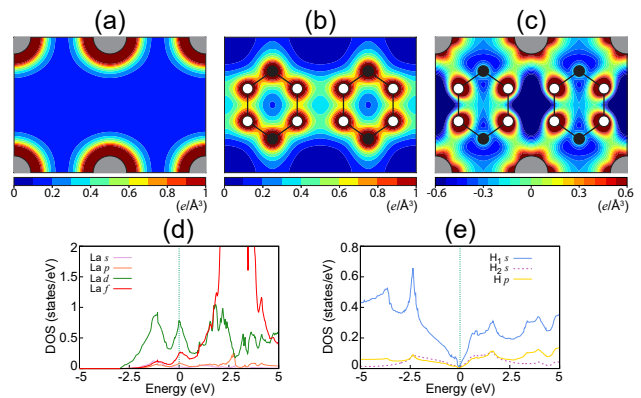}
\caption{(Color online) Calculated charge densities of (a) ${\rho}_{\rm La}$ and (b) ${\rho}_{\rm H}$ in the LaH$_0$ lattice and the La$_0$H$_{10}$ lattice, respectively. The charge density difference ${\Delta}{\rho}$ (defined in the text) is displayed in (c). The corresponding partial DOS of the LaH$_0$ and La$_0$H$_{10}$ lattices are given in (d) and (e), respectively.}
\end{figure}

It has been known that the existence of vHs near $E_{\rm F}$ plays an important role in the BCS-type superconductivity~\cite{Vanhove,Vanhove1,Vanhove2}. To investigate how the electronic states around the vHs points are strongly coupled to phonon modes, we calculate the phononic properties of LaH$_{10}$. Figure 4(a) shows the calculated phonon spectrum with the EPC strength, the phonon DOS projected onto selected atoms, and the Eliashberg function ${\alpha}^{2}F({\omega})$ with the integrated EPC constant ${\lambda}({\omega})$. Our results for the phonon dispersion and EPC strength are in good agreement with those obtained by a previous DFT calculation~\cite{Hydride5}. Interestingly, the phonon DOS is found to be divided into the three regions I, II, and III [see Fig. 4(a)]. Here, the region I is mostly contributed by the vibrations of La atoms, the region II is due to a nearly equal mixture of vibrations from H$_1$ and H$_2$ atoms, and the region III arises mainly from the vibrations of H$_1$ atoms. Therefore, we estimate that the optical phonon modes of H$_1$ atoms contribute to ${\sim}$63\% of ${\lambda}({\omega})$, while the acoustic (optical) phonon modes of La (H$_2$) atoms contribute to ${\sim}$15 (22)\% of ${\lambda}({\omega})$. This estimation indicates that H$_1$-derived high frequency optical phonon modes comprise substantial contributions to high $T_{\rm c}$. We note that the frequencies of whole optical modes range between ${\sim}$710 and ${\sim}$2350 cm$^{-1}$, which are relatively higher than those (500 ${\sim}$ 2000 cm$^{-1}$) of compressed H$_3$S~\cite{Duan}. By numerically solving the Eliashberg equations~\cite{Eliash} with the typical Coulomb pseudopotential parameter of ${\mu}^*$ = 0.13~\cite{Hydride5,Hydride7}, we estimate $T_{\rm c}$ ${\approx}$ 335 K (see Fig. S7 of the Supplemental Material~\cite{SM}), which is much larger than that (${\sim}$270 K) obtained using the Allen-Dynes~\cite{Allen}. However, in order to choose a proper value of ${\mu}^*$ that fits the experimental data of $T_{\rm c}$, we calculate the dependence of $T_{\rm c}$ on ${\mu}^*$ using the Eliashberg equations~\cite{Eliash}. The results are given in Fig. S7 of the Supplemental Material~\cite{SM}. We find that  ${\mu}^*$ = 0.38 gives $T_{\rm c}$ ${\approx}$ 220 K at 300 GPa, which can be interpolated to reach the experimental value of  $T_{\rm c}$ = 260 K (measured at 190 GPa~\cite{LaH2}) using the theoretical result (see Fig. 7 of Ref. ~\cite{Hydride5}) of $T_{\rm c}$ versus pressure. Considering that LaH$_{10}$ and H$_3$S have similar vibrational and electronic features such as the range of high phonon frequencies and the existence of a double-shaped vHs near $E_{\rm f}$, we expect that for LaH$_{10}$, nonadiabatic effects with the lowest-order vertex corrections would reduce ${\mu}^*$ by ${\sim}$10${\%}$ compared to the value obtained using the adiabatic approximation, as previously estimated from H$_3$S~\cite{Scirep}. In Fig. 4(a), the size of the circles on the phonon dispersion represents the EPC strength, which indicates that some strong EPCs occur in the ${\Gamma}$, $X$, $L$, and $K$ points. The corresponding phonon wave vectors may be associated with the Fermi-surface nesting. In Fig. 4(b), we plot the Fermi surface with the major nesting vectors $q_{\Gamma}$, $q_{X}$, $q_{L}$, and $q_{K}$. Here, $q_{\Gamma}$ and $q_{X}$ match with the spanning vectors between the vHs points around the $L$ points. For $q_{L}$ and $q_{K}$, the electronic states at the vHs points are likely coupled to the ${\Gamma}$-centered, polyhedron-shaped Fermi surface and the flat Fermi-surface portions, respectively, participating in EPC [see Fig. 4(b)]. Therefore, we can say that the coupling vectors between the vHs points and the Fermi surface account for the strong EPCs in the ${\Gamma}$, $X$, $L$, and $K$ points.

\begin{figure}[htb]
\includegraphics[width=8.0cm]{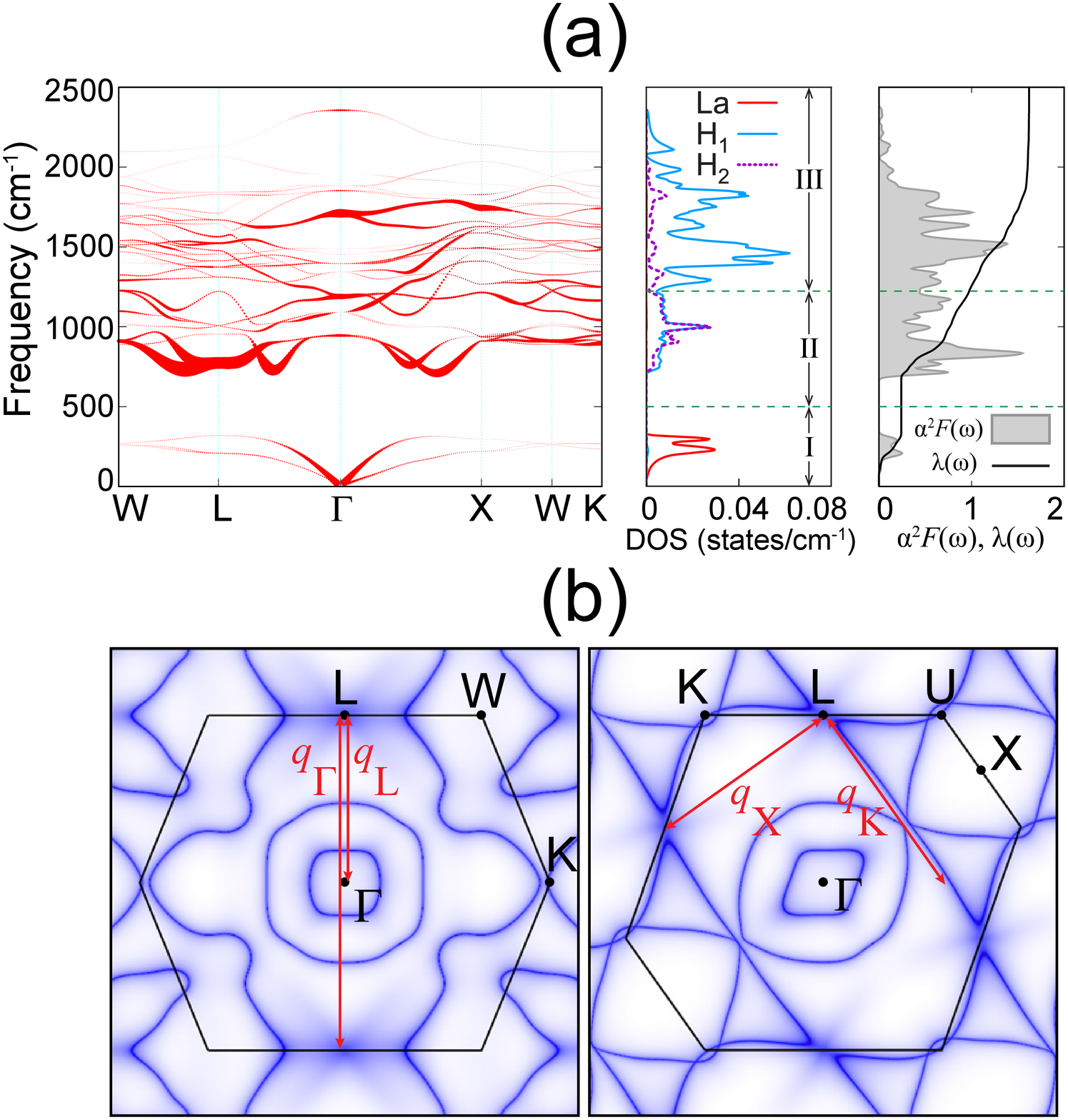}
\caption{(Color online) (a) Calculated phonon spectrum, phonon DOS projected onto selected atoms, Eliashberg function ${\alpha}^{2}F({\omega})$, and integrated EPC constant ${\lambda}({\omega})$ of LaH$_{10}$. The size of circles on the phonon dispersion is proportional to the EPC strength. In (b), Fermi surface obtained using WannierTools~\cite{wanniertool} is displayed on two different 2D cross sections of the Brillouin zone. Here, the coupling vectors of the vHs points and the Fermi surface are indicated by the double-headed arrows.}
\end{figure}

To conclude, our first-principles calculations for the electronic, bonding, and phononic properties of the compressed $fcc$-LaH$_{10}$ phase have revealed two unique features that play important roles in increasing $T_{\rm c}$ of the BCS-type SC. One is the existence of vHs near $E_{\rm F}$, which originates from the hole-like and electron-like bands arising from the topological DNL states, and the other represents H-derived high vibrational frequencies that are induced via the unusual, intricate bonding network with anionic La and both cationic and anionic H atoms. These two features cooperate to produce a strong EPC, thereby enhancing $T_{\rm c}$ in $fcc$-LaH$_{10}$. It is noted that the $fcc$-LaH$_{10}$ phase can be stabilized at higher pressures above ${\sim}$170 GPa~\cite{LaH}, while it is transformed into a lower symmetry $C2/m$-LaH$_{10}$ phase on decompression below 170 GPa~\cite{LaHdistort}. Because of the disappearance of vHs near $E_{\rm F}$ in the $C2/m$-LaH$_{10}$ phase~\cite{LaHdistort}, its $T_{\rm c}$ is expected to be lower than that of the $fcc$-LaH$_{10}$ phase. As a matter of fact, recent experiments reported that $T_{\rm c}$ was observed to be ${\sim}$215 K at ${\sim}$150 GPa~\cite{LaH4}, which is lower than ${\sim}$260 (250) K at ${\sim}$190 (170) GPa~\cite{LaH2, LaH3}. The present findings not only provided an explanation for the microscopic origin of the observed high-$T_{\rm c}$ SC in compressed $fcc$-LaH$_{10}$, but also will stimulate further research to explore prospective room-temperature topological superconductors~\cite{Majo1,Majo3} in highly compressed hydrides.

\vspace{0.4cm}

\noindent {\bf Acknowledgement.}
We are grateful to Prof. Yanming Ma for his introduction of the present topic. We also thank Prof. Y. Jia for his support. This work was supported by the National Research Foundation of Korea (NRF) grant funded by the Korean Government (Grant Nos. 2016K1A4A3914691 and 2015M3D1A1070609). The calculations were performed by the KISTI Supercomputing Center through the Strategic Support Program (Program No. KSC-2017-C3-0080) for the supercomputing application research and by the High Performance Computational Center of Henan University.  \\

L. L., C. W., and S. Y. contributed equally to this work.

                       %%%%%  REFERENCES  %%%%%
\noindent $^{*}$ Corresponding author: chojh@hanyang.ac.kr

\newpage
\onecolumngrid
%\appendix
%\titleformat*{\section}{\Large\bfseries}
\titleformat*{\section}{\LARGE\bfseries}

\renewcommand{\thefigure}{S\arabic{figure}}
\setcounter{figure}{0}

\vspace{1.2cm}

\section{Supplemental Material for "Origin of High-Temperature Superconductivity in LaH$_{10}$"}
%\vspace{0.2cm}
\begin{flushleft}
{\bf 1. Band projections onto the H and La orbitals}
\begin{figure}[ht]
\includegraphics[width=10cm]{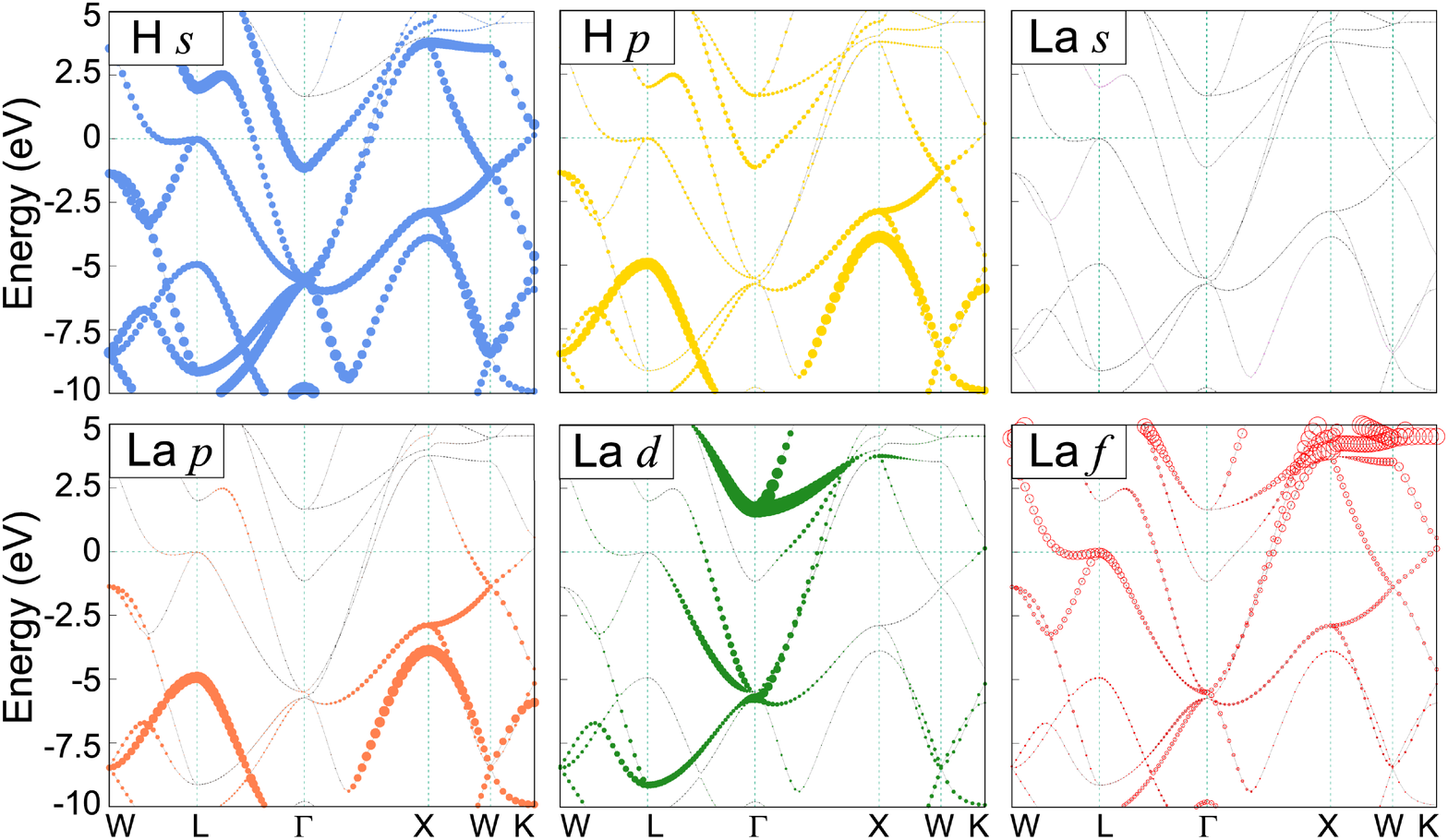}
\caption{ Calculated bands projected onto the H $s$, H $p$, La $s$, La $p$, and La $f$ orbitals in LaH$_{10}$. Here, the radii of circles are proportional to the weights of the corresponding orbitals. The H $s$ and La $f$ orbitals are more dominant components of the electronic state located near $E_{\rm F}$, compared to other orbitals.}
\end{figure}

\vspace{0.4cm}

%\newpage
{\bf 2.  Calculated band structure, partial density of states, DNLs, and Berry curvature in H$_3$S}
\begin{figure}[hb]
\includegraphics[width=11cm]{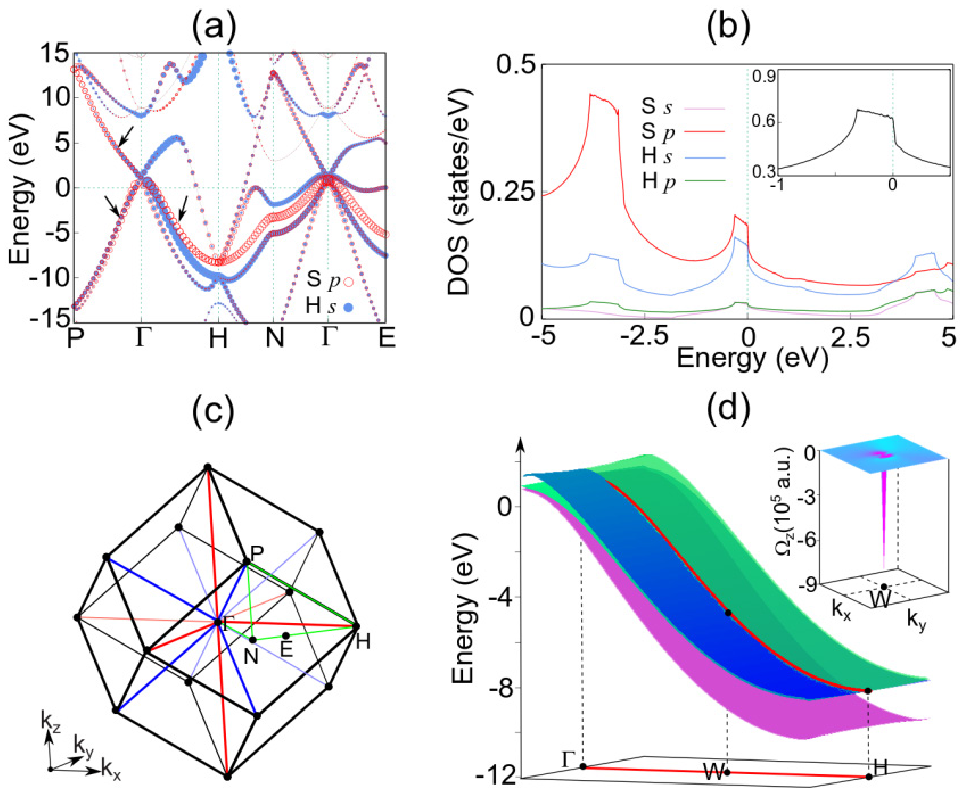}
\caption{  (a) Calculated band structure and (b) partial DOS of H$_3$S. In (a), the bands projected onto the H $s$ and S $p$ orbitals are displayed with circles whose radii are proportional to the weights of the corresponding orbitals. The arrows in (a) indicate the DNLs along the $\overline{{\Gamma}H}$ and $\overline{{\Gamma}P}$ lines. The energy zero represents $E_{\rm F}$. A close-up of the total DOS around the vHs is given in the inset of (b). In (c), all the DNLs along the symmetry-equivalent $\overline{{\Gamma}H}$ and $\overline{{\Gamma}P}$ lines are drawn in the first Brillouin zone. The 1D dispersion of the DNL bands along the $\overline{{\Gamma}H}$ line is displayed in (d), together with the distribution of a Berry curvature component ${\Omega}_z$ around a band touching point W in the DNL.}
\end{figure}
\newpage
{\bf 3. Comparison of the electronic bands obtained using the DFT and tight-binding Hamiltonian calculations}
\begin{figure}[ht]
\includegraphics[width=8cm]{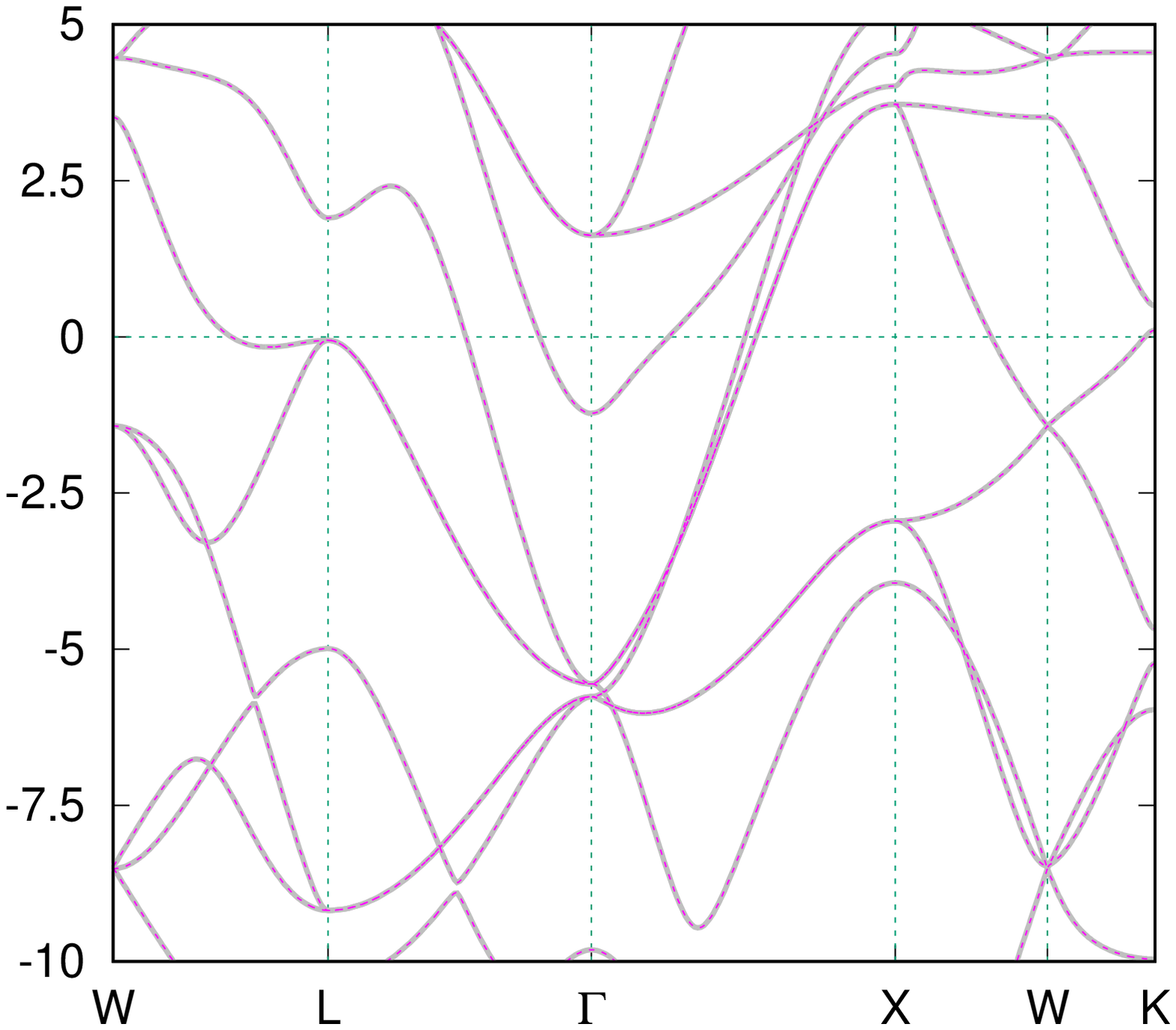}
\caption{ Band structure of LaH$_{10}$ obtained using the tight-binding Hamiltonian with maximally localized Wannier functions (MLWF), in comparison with that obtained using the DFT calculation. The DFT band structure is well reproduced by using the Wannier90 package [1]. Here, the wave functions are projected onto the twenty-six H $s$, La $s$, La $p$, La $d$, and La $f$ orbitals.}
\end{figure}

{\bf 4. Band structure of LaH$_{10}$ obtained using the DFT calculation with including the SOC}
\begin{figure}[hb]
\includegraphics[width=16cm]{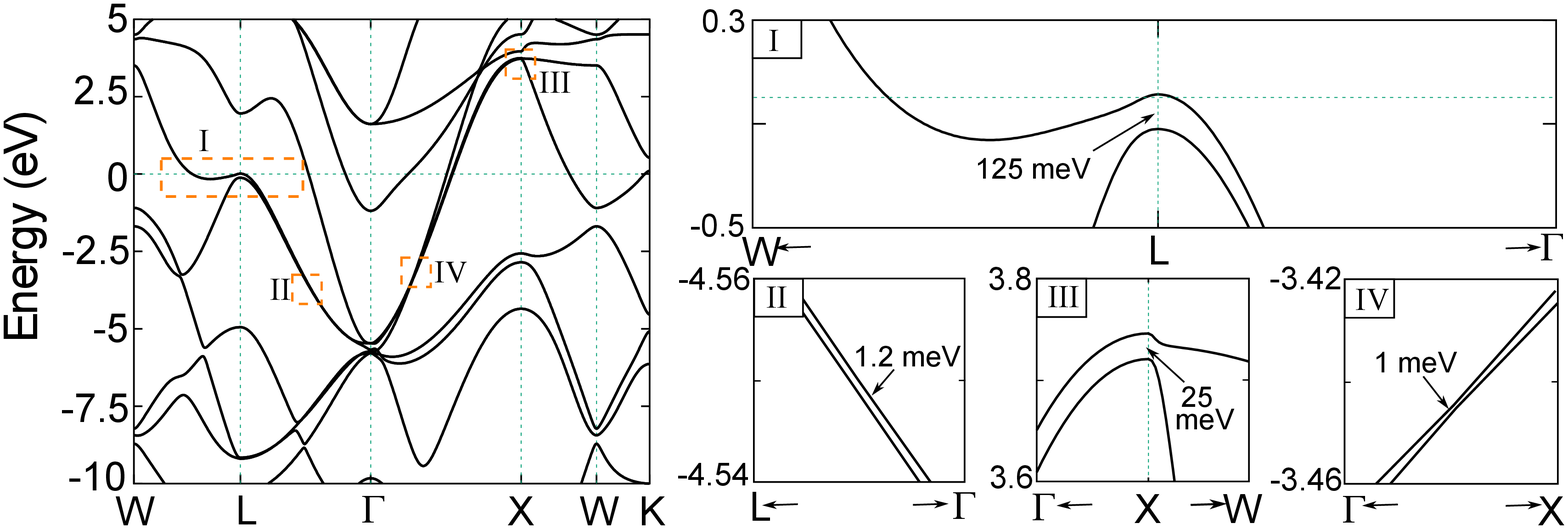}
\caption{ Band structure of LaH$_{10}$ obtained using the DFT+SOC calculation. Zoom-in bands in the dashed squares I, II, III, and IV exhibit the SOC-induced gaps.}
\end{figure}

\vspace{0.4cm}
\newpage

{\bf 5. Total DOS obtained using the DFT calculation with including the SOC}
\begin{figure}[ht]
\includegraphics[width=8cm]{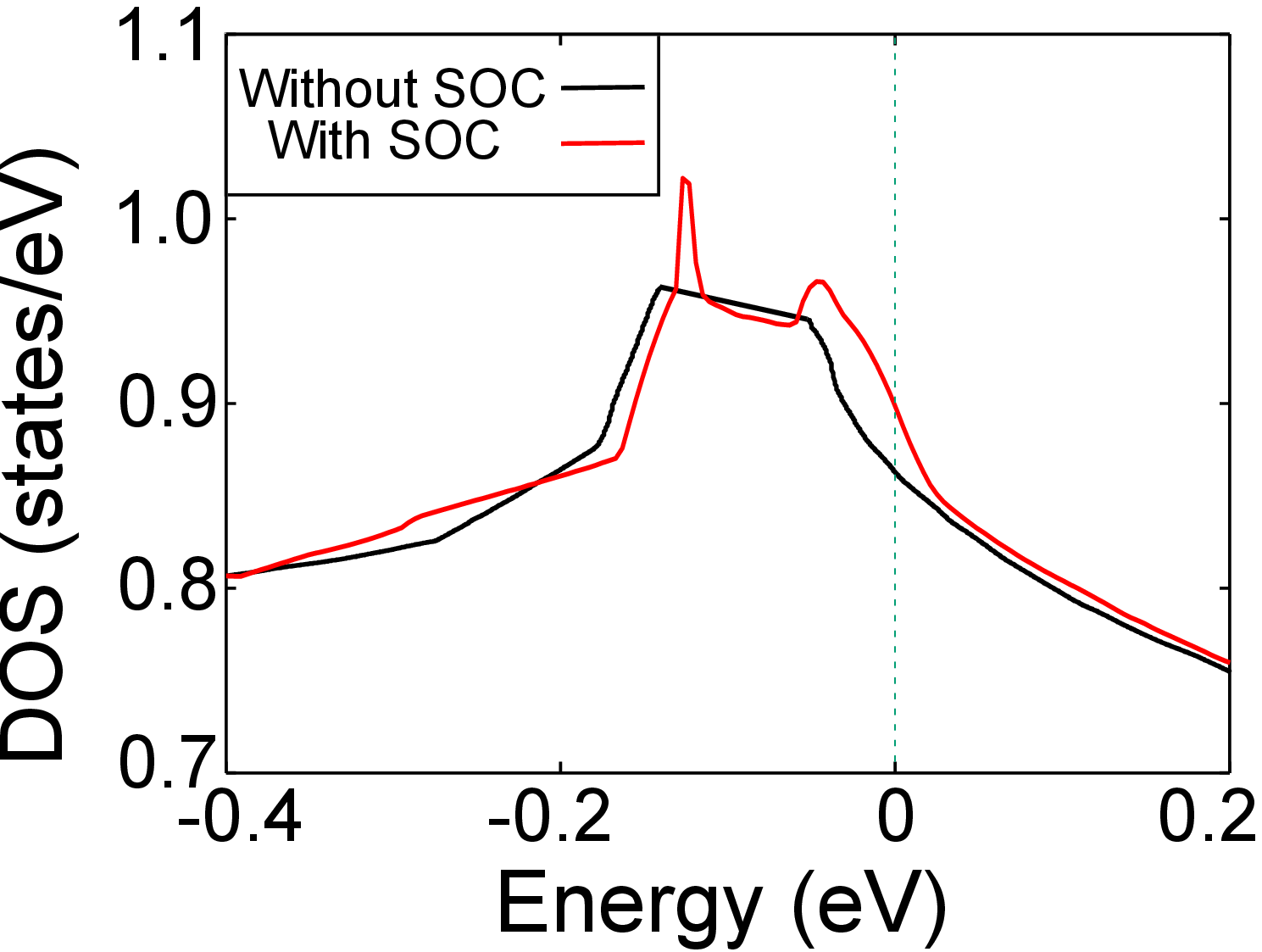}
\caption{ Total DOS obtained using the DFT+SOC calculation. For comparison, the total DOS obtained without the inclusion of SOC is also given.}
\end{figure}
\vspace{0.4cm}

{\bf 6. Frequencies of the ${\Gamma}$-phonon modes of LaH$_{10}$ and La$_0$H$_{10}$}
\begin{figure}[hb]
\includegraphics[angle=270,origin=c,width=8cm]{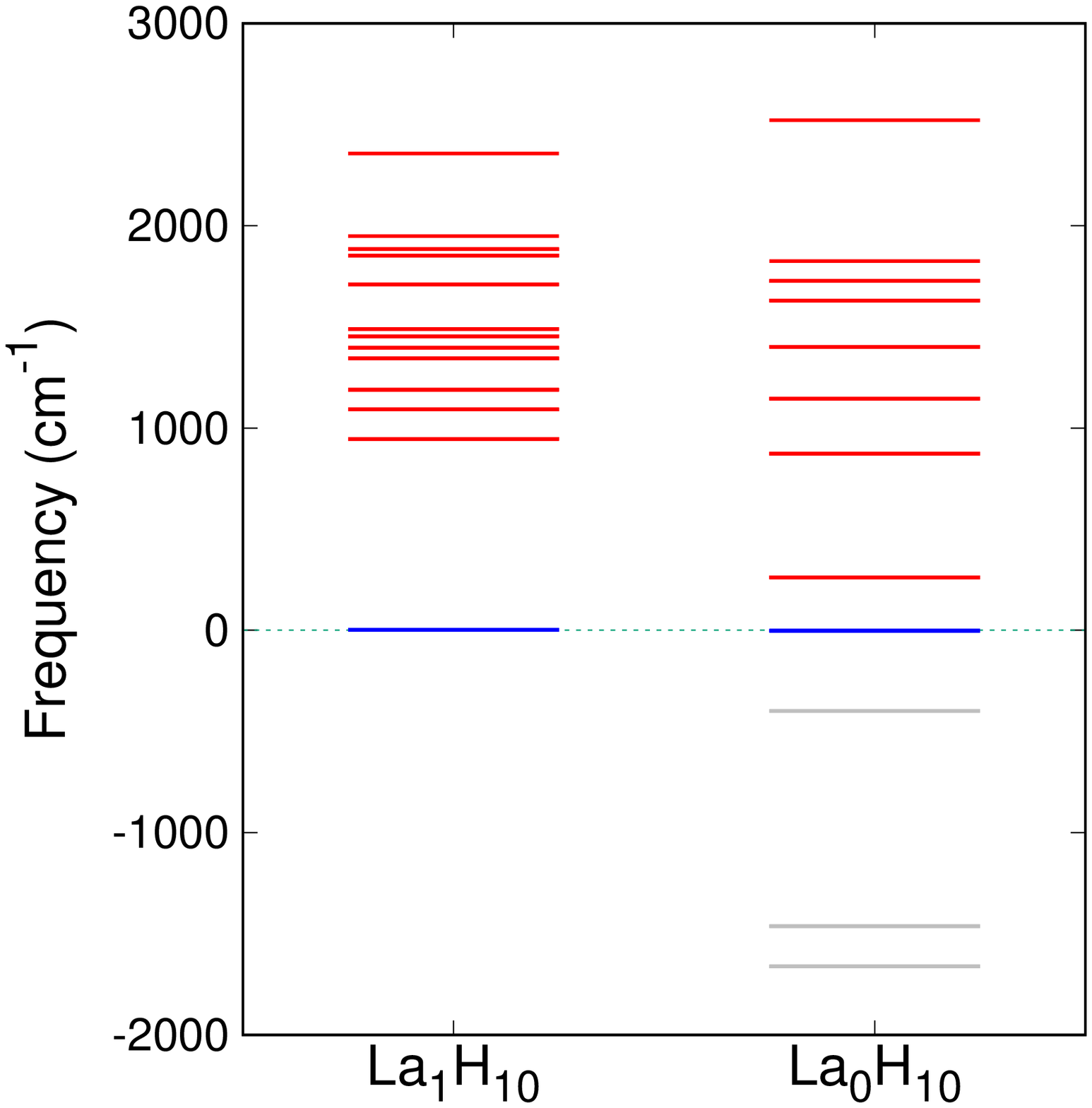}
\caption{ Calculated frequencies of the ${\Gamma}$-phonon modes of LaH$_{10}$ and La$_0$H$_{10}$ (LaH$_{10}$ structure without La atoms). The frequencies of optical modes are much reduced in La$_0$H$_{10}$, compared to those in LaH$_{10}$}
\end{figure}
\newpage
{\bf 7. Superconducting energy gap and transition temperature}
\begin{figure}[htb]
\includegraphics[width=12cm]{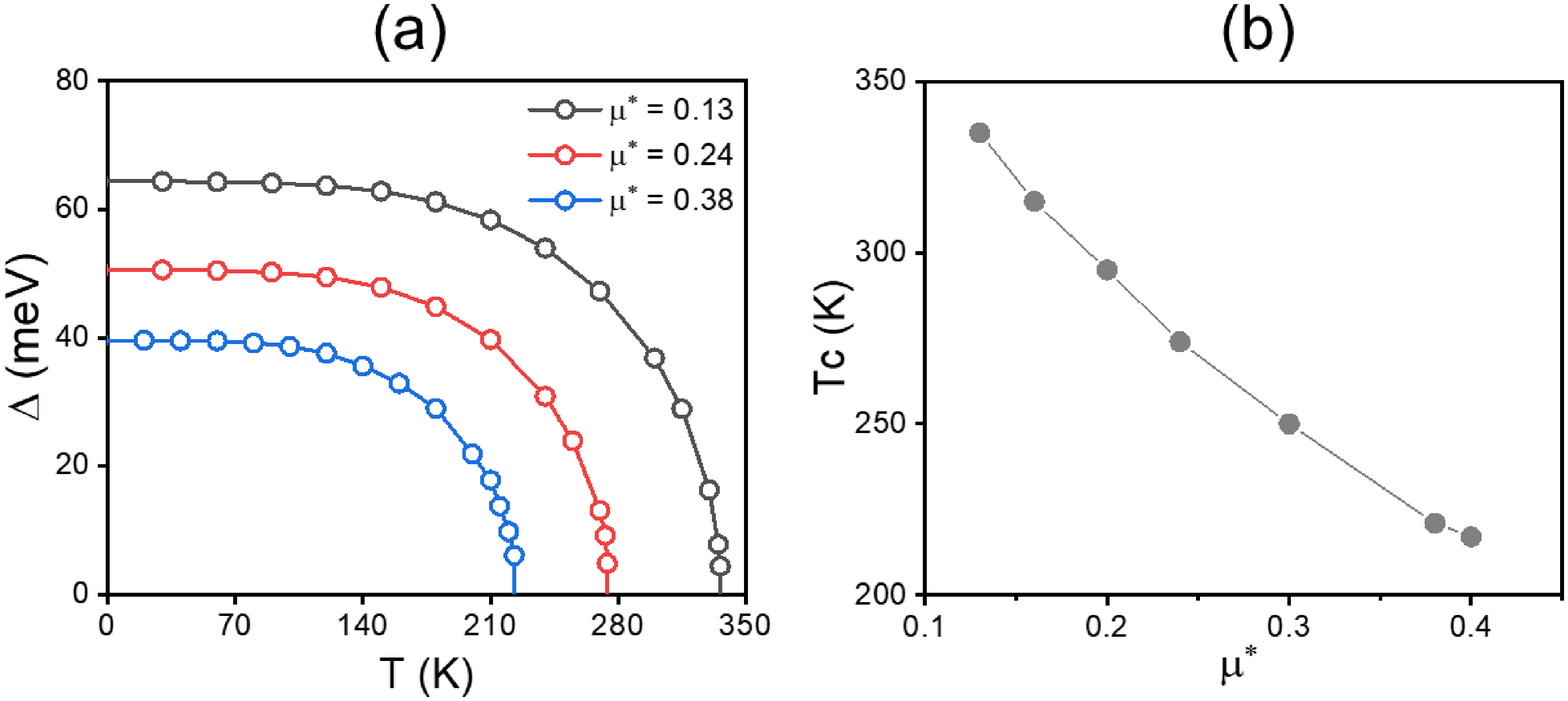}
\caption{ (a) Calculated superconducting energy gap ${\Delta}$ as a function of temperature $T$ and (b) superconducting transition temperature $T_{\rm c}$ as a function of ${\mu}^*$.}
\end{figure}

\end{flushleft}

\end{document}